\documentclass[article, reprint, superscriptaddress, showpacs, showkeys, prd,14pt]{revtex4-1}

\usepackage{amsmath} 
\usepackage{graphicx} 
\usepackage{verbatim} 
\usepackage{color} 
\usepackage{subfigure} 
\usepackage{hyperref} 
\usepackage{longtable}
\usepackage{dcolumn}

\begin{document}

\title{Hadrons to Vector Mesons - Photons \\  and the Electromagnetic Form Factors}

\author{Vlasios Petousis}  \email{petousis@ucy.ac.cy} 
\affiliation{Department of Physics, University of Cyprus, Nicosia, CY-1678, Cyprus}
\date{\today}

\begin{abstract}
In Ultra Relativistic Heavy-Ion Collisions (URHIC's) the light vector mesons play a crucial role in the description of hadronic interactions in a non-pertrubative QCD region. In this region effective hadron models use composite hadrons and mesons as field carriers instead of quarks and gluons. This theoretical review article tries to give a short way of understanding the underlying theory that we have for the conversion of vector mesons into photons together with the so-called Vector Dominance Model (VDM) and the electromagnetic form factors.

\end{abstract}

\keywords{Vector Dominance Model, Electromagnetic Form Factors}
\maketitle
\section{Introduction}
The last years a lot of experimental and theoretical efforts have been made to investigate the physics behind of  the  Ultra Relativistic Heavy-Ion Collisions (URHIC's). Today the efforts towards the deconfinement phase of the hadrons and the research for the chiral symmetry restoration phase have become more systematic. In this research, tools like, light vector mesons production, search for strangeness and di-leptonic production are in the front line of the battle. Each tool has its own advantages and disadvantages but one is for sure, that all this years a great progress in understanding the physics behind have been made. In the following two sections we summarize the work depend on hadron structure of photon and the Vector Dominance Model (VDM) and also the Electromagnetic Form Factors. Very important and useful tools to the understanding of the nature of the Quark Gluon Plasma formation at the Ultra Relativistic Heavy-Ion Collisions. The vector mesons $\rho$, $\omega$ and $\phi$ play an important role in the description of hadronic interactions in a non-pertrurbative region. In this region effective hadron models use composite hadrons and mesons as field carriers instead of quarks and gluons. In these models, vector mesons are introduced together with photons as a gauge bosons of an implicit gauge symmetry in a close analogy to spontaneously broken $SU(2)\times U(1)$ symmetry of the electro-weak interactions. 
The fundamental role of the vector mesons was pointed out first by Sakurai in the Vector Dominance Model (VDM)\cite{1} who introduced the hadron-vector meson interactions by analogy with the photon-electron interactions in Quantum Electrodynamics (QED). According to this model, the coupling of the hadrons to vector mesons is described by an universal coupling constant which conserved in all hadronic interactions. Also the same coupling constant and the vector meson mass describes the vector meson conversion into a photon. 
\section{Vector Meson Conversion into Photon - Vector Dominance Model}
In field theory, the particle interactions are described via Lagrangians of interacting current fields. For example, in the QED Lagrangian the quark-photon electromagnetic interaction is described by a product of the quark electromagnetic current ${j^{EM}_{\mu}=Q\overline{\psi}\gamma_{\mu}\psi}$ with photon vector field ${A_{\mu}}$: ${L^{EM}_{\mu}=ej^{EM}_{\mu}A^{\mu}}$. The Vector meson Dominance Model (VDM) assumes that at low energies where hadrons rather than quarks are appropriate degrees of freedom, the quark electromagnetic current is carried entirely by a vector meson fields. The typical energy range where this hypothesis holds is given by a four-momentum of the exchanged photon of ${q^{2}\prec1}$ ${Ge{{V}^{2}}}$. This assumption results in so-called field current, valid exactly at the vector meson poles:
\begin{equation}
{j^{EM}=\sum\gamma_{V}\Psi^{V}=\sum_{V}\frac{m^{2}_{V}}{f_{V}}\Psi^{V}}\label{eq:1.1}
\end{equation}
where ${\Psi^{V}}$ are the vector meson fields and ${\gamma_{V}}$ is the conversion factor given by the vector meson mass $m_{V}$ and the universal coupling constant ${f_{V}}$. A connection between the vector fields ${\Psi^{V}}$ and the quark electromagnetic current $j^{EM}$ is easily obtained in the framework of the second quantization. In this framework, the $j^{EM}$ is a creation operator acting on the QCD vacuum and creating quark-antiquark pairs witch overlap with the mesonic states. 
The coupling constant ${f_{V}}$ can be obtained from the measured di-electron vector meson decay widths. For example the matrix element for the ${\rho^{0}\rightarrow e^{+} + e^{-}}$ decay, the ${\left| M (V e^{+}e^{-}) \right|}$ can be calculated within the VDM using the graph in Fig.1B and the following factorization: 
 \begin{equation}
\begin{split}
{\left| M (V e^{+}e^{-}) \right|^{2}= \left| M (V\gamma^{*}) \right|^{2}\frac{1}{q^{2}}\left| M (  \gamma^{*} e^{+}e^{-}) \right|^{2}} \\ {=\frac{e^{2}m^{4}_{V}}{f^{2}_{V}}\frac{1}{m^{4}_{V}}\left| M (\gamma^{*} e^{+}e^{-}) \right|^{2}}\label{eq:1.2}
\end{split}
\end{equation}
The last matrix element describes the conversion of the virtual ${\gamma^{*}}$ with the four-vector ${q^{2}=m^{2}_{V}}$ into a positron-electron pair and is given by:
\begin{equation}
{\left|M(\gamma^{*}e^{+}e^{-}) \right|^{2}=\frac{4\pi\alpha}{3}q^{2}}\label{eq:1.3} 
\end{equation}
 \begin{equation}
{\left|M(Ve^{+}e^{-}) \right|^{2}=\frac{16\pi^{2}\alpha^{2}}{3f^{2}_{V}}m^{2}_{V}}\label{eq:1.4} 
\end{equation}
Using eq.(\ref{eq:1.3}) and neglecting the electron mass, one gets the di-electron vector meson decay width\cite{1}:
\begin{equation}
{\Gamma(Ae^{+}e^{-})=\frac{p_{CM}}{8\pi m^{2}_{A}}\left|  M \right|^{2}}\label{eq:1.5} 
\end{equation}    
\begin{equation}
{\Gamma^{e^{+}e^{-}}_{V} =4\frac{\left|  p_{cm} \right|}{8\pi m^{2}_{V}}\left|  M_{Ve^{+}e^{-}}  \right|^{2}=\frac{4\pi\alpha^{2}}{3f^{2}_{V}}m_{V}}\label{eq:1.6} 
\end{equation}
where ${{p_{cm}=\frac{m_{V}}{2}}}$ is the momentum of the daughter electron in the center mass frame.
From the measured di-electron experimental decay widths one can therefore obtain the universal coupling constants for the light vector mesons:
\begin{equation}
{{f_{\rho}=5.03} \hspace{5 mm} {f_{\omega}=17.05} \hspace{5 mm}{f_{\phi}=-12.89}}\label{eq:1.7} 
\end{equation}
and as one can see from eq.(\ref{eq:1.6}) that the ratios ${\Gamma^{e^{+}e^{-}}_{V}}$ for the $\rho$, $\omega$ and $\phi$ mesons is proportional to $\frac{m_{V}}{f^{2}_{V}}$. They also agree with the ratios of the squared numerical coefficients resulting from the $SU(3)$ decomposition, namely:
\begin{figure}[h]
\centering
\includegraphics[width=70mm, height=40mm]{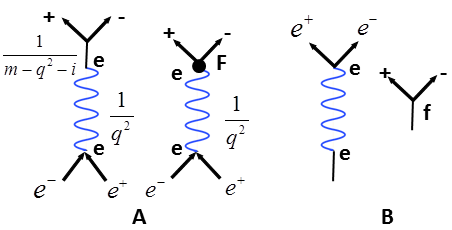}
\caption{A: The Feynman graphs (time flows from left to right) for the ${e^{+}e^{-}}$ annihilation with ${\pi^{+} \pi^{-}}$ emission. According to VDM hypothesis a time-like photon (${{q}^{2}}\succ 0$)  couples to a pion via an intermediate vector $\rho^{0}$ meson state. B: Vertex coupling constants ${\gamma }_{\rho }$ and $f_{\rho\pi\pi}$ can be derived from the $\rho \to {{e}^{+}}{{e}^{-}}$ and $\rho \to {{\pi }^{+}}{{\pi }^{-}}$ decays, respectively.}
\end{figure}
\begin{equation}
\begin{split}
{\Gamma^{e^{+}e^{-}}_{\rho} : \Gamma^{e^{+}e^{-}}_{\omega} : \Gamma^{e^{+}e^{-}}_{\phi} =} \\ {\frac{m_{\rho}}{f^{2}_{V}} :\frac{m_{\omega}}{9f^{2}_{V}}:\frac{m_{\phi}}{4.5f^{2}_{V}}\approx \frac{1}{2}:\frac{1}{18}:\frac{1}{9}}\label{eq:1.8}  
\end{split}
\end{equation} 
where $f_{V}=f_{\rho}=5.7$ as value selected. Exception is the $\rho$ meson for which one observes a larger discrepancy. The deviation on $f_{\rho}$ from the ideal $SU(3)$ value $f_{V}$ can be explained by a large contribution on the one-pion loop to the $\rho$-self energy at the effective Lagrangian approach \cite{2}.
Another part of the VDM hypothesis is the assumption that the same coupling constant $f_{V}$ describes also the meson-hadron interaction and therefore connects strong and electromagnetic interactions by the vector mesons gauge fields. The interaction Lagrangian can be written by analogy to the electromagnetic interaction introduced above:
\begin{equation}
{L^{H}_{int}=f_{V}j^{H}_{\mu}\Psi^{\mu}}\label{eq:1.9}  
\end{equation}
where ${j^{H}_{\mu}}$ is the hadronic vector current and ${\Psi^{\mu}}$ is the vector mason field (here we replaced the ${V}$ index form eq.(\ref{eq:1.1}) with ${\mu}$). The coupling constant $f_{V}$ can be interpreted as a conserved hadronic charge of the vector meson and for example should be the same for the  ${\rho-NN}$ and ${\rho-\pi\pi}$ couplings and equal to $f_{V}=5.7$ derived from the $\rho$ meson di-electron decay. This can be verified by the analysis of the di-pion decay width of the $\rho$ meson. The $\Gamma_{\rho}^{\pi^{+}\pi^{-}}$ decay width can be calculated in a similar way as for di-electron decay using the graph of the Fig.1B. The matrix elements is given by\cite{3}:
\begin{equation}
{\left|  M(\pi^{+}\pi^{-}\rho^{0}) \right|^{2}=4f^{2}_{\rho\pi\pi}p^{2}_{cm}}\label{eq:1.10}  
\end{equation}  
and finally we get:
\begin{equation}
{\Gamma(\rho\pi^{+}\pi^{-})=\frac{2}{3}\frac{\left| p_{cm} \right|^{3}}{m^{2}_{p}}(\frac{f^{2}_{\rho\pi\pi}}{4\pi})=151MeV}\label{eq:1.11}  
\end{equation}
where 
\begin{equation}
{\left| p_{cm} \right|=\sqrt[]{(\frac{m_{\rho}}{2})^2-m^{2}_{\pi}}}\label{eq:1.12}
\end{equation}
Using the experimental value from eq.(\ref{eq:1.11}) we obtain that ${f_{\rho\pi\pi}}=6.05$ which is close to the universal value ${f_{V}=5.7}$.
Should be mentioned here an important aspect of broad resonances such as the $\rho$ meson. The decay widths in eq.(\ref{eq:1.5}) and eq.(\ref{eq:1.11}) are valid strictly for the meson poles $m_{R}$. For the broad $\rho^{0}$, the respective decay widths are mass dependent and for the di-electron decay as given in\cite{4},\cite{5}is:
\begin{equation}
{\Gamma^{e^{+}e^{-}}_{\rho}(M)=\Gamma^{e^{+}e^{-}}_{\rho}(m_{R})(\frac{m_{R}}{M})^{3}}\label{eq:1.13}
\end{equation}  
for the di-pion decay is:
\begin{equation}
{\Gamma^{\pi^{+}\pi^{-}}_{\rho}(M)=\Gamma^{\pi^{+}\pi^{-}}_{\rho}(m_{R})(\frac{m_{R}}{M})(\frac{p_{cm}(M)}{p_{cm}(m_{R})})^{3}}\label{eq:1.14}
\end{equation}
An important consequence of eq.(\ref{eq:1.14}) is that in contrary to the di-pion decay the di-electron decay yield increases at low masses.

\section{The Electromagnetic Form Factors of Hadrons}
The dominance of the vector masons in the hadron-photon interaction vertex which discussed at the previous paragraph is conspicuous in the  electromagnetic form factors of the hadrons. Let us consider the electron-positron annihilation with a pion production. The cross section ratio of the positron annihilation with the production of even number of pions, normalized to the cross section for the di-muon production as\cite{2}:
\begin{equation}
{R=\frac{\sigma^{n\pi}_{e^{+}e^{-}}}{\sigma^{\mu^{+}\mu^{-}}_{e^{+}e^{-}}}}\label{eq:2.1}
\end{equation}  
At large $Q^{2}$ the observed $R$ values is one of the proofs for the quark-hadron structure. As predicted by the Quark Parton Model and QED, for large four-momentum transfers (${q^{2}\succ1}$ $GeV^{2}$) the $R$ value is proportional to the product of the electromagnetic current, carried by electrons and quarks according to:
\begin{equation}
{R=({j^{EM}_e}\frac{1}{q^{2}_{quarks}})^{2}}\label{eq:2.2}
\end{equation}
The result of the leading order QCD calculation, shows that $R$ is equal to the sum of squared light quarks charges $Q_{q}$ multiplied by the number of colors $N_{c}$, ($N_{c}=3$):
\begin{equation}
\begin{split}
{R=N_{c}\sum_{u,d,s}Q_{q}^{2}=}\\{3[(\frac{2}{3})^{2}+(\frac{-1}{3})^{2}+(\frac{1}{3})^{2}](1+\frac{\alpha_{s}(q^{2})}{\pi})}\label{eq:2.3}
\end{split}
\end{equation}
where ${\alpha_{s}(q^{2})}$ is the QCD "running" coupling constant.
At low energies (${q^{2}\prec1}$ $GeV^{2}$) according to VDM the quark electromagnetic current is carried out by the vector mesons what has been confirmed by the annihilation experiments. The relation of the quark electromagnetic current $j^{EM}_{\mu}$ and the pion electromagnetic form factor $F_{\pi}(q^{2})$ is given by the following matrix element\cite{2}:
\begin{equation}
{\left\langle \pi^{\pm} \left| j^{EM}_{\mu}(0) \right| \pi^{\pm}\right\rangle=\pm(k+k^{'})_{\mu}F_{\pi}(q^{2})}\label{eq:2.4}
\end{equation}
where $q^{2}$ is the four-momentum transfer ${q^{2}=(k-k^{'})^{2}}$ of scattered particle. The form factor can be obtained in a time-like region $(q^{2}\succ0)$ from the annihilation experiments discussed above. Here we consider the cross section for an inverse reaction of pion annihilation ${\sigma^{e^{+}e^{-}}_{\pi^{+}\pi^{-}}}$ important for di-electron production. The cross section for this reaction can be calculated according to the production graph shown in Fig.1A assuming the VDM. The reaction transition matrix element can be constructed connecting the matrix elements for the $\rho^{0}\rightarrow e^{+}e^{-}$ from eq.(\ref{eq:1.5}) and for $\rho^{0}\rightarrow \pi^{+}\pi^{-}$ from eq.(\ref{eq:1.11}) decays, using the $\rho^{0}$ propagator: 
\begin{equation}
\begin{split}
{\left|  M(\pi^{+}\pi^{-}e^{+}e^{-}) \right|^{2}=}\\ {\left|  M(\pi^{+}\pi^{-}\rho^{0}) \right|^{2}\frac{\left| M(\rho^{0}e^{+}e^{-}) \right|^{2}}{(m^{2}_{\rho}-M^{2})-m^{2}_{\rho}\Gamma^{2}_{\rho}}}\label{eq:2.5}
\end{split}
\end{equation}
Using the formula for the production cross section for a two-body scattering process in the center mass frame\cite{6}one gets:
\begin{equation}
{\sigma^{e^{+}e^{-}}_{\pi^{+}\pi^{-}}(M)=\frac{1}{4\pi\sqrt[]{s}}\frac{p_{f}}{p_{i}}\left| M(\pi^{+}\pi^{-}e^{+}e^{-}) \right|^{2}}\label{eq:2.6}
\end{equation}
\begin{equation}
{\sigma^{e^{+}e^{-}}_{\pi^{+}\pi^{-}}(M)=\frac{4\pi\alpha^{2}}{3M^{3}}\sqrt[]{M^{2}-4m^{2}_\pi}\left| F_{\pi} \right|^{2}}\label{eq:2.7}
\end{equation}
where $p_{i}$ and $p_{f}$ is the initial and final state momenta for pion and electron respectively, calculated in the center mass frame also. The time-like pion electromagnetic form factor $F_{\pi}$ in eq.(\ref{eq:2.7}) contains the $\rho$ meson propagator and the ${\gamma_{\rho}}$, ${f_{\rho\pi\pi}}$ coupling constants. It can be also easily calculated comparing the internal lines of two production graphs in Fig.1A, where $F_{\pi}$ is shown as a blob in the right graph:
\begin{equation}
{\left|F_{\pi}(M)\right|^{2}=\frac{f^{2}_{\rho\pi\pi}}{f^{2}_{V}}\frac{m^{4}_{\rho}}{(M^{2}-m^{2}_{\rho})^{2}+M^{2}_{\rho}\Gamma^{2}_{\rho}}}\label{eq:2.8}
\end{equation}
The precise description of ${F_{\pi}(q^{2})}$, obtained by the eq.(\ref{eq:2.8})  including $\rho$-$\omega$ mixing and one-loop pion contribution\cite{2},\cite{7}. Time-like electromagnetic form factors for other hadrons can be obtained by studying invariant mass distributions of di-electrons streaming from the Dalitz $A\rightarrow Xe^{+}e^{-}$ decays. According to the VDM, these decays proceed as shown in Fig.2, as two sequential body decays: $A\rightarrow X\gamma^{*}(M)$ with the sub sequential virtual $\gamma^{*}\rightarrow Xe^{+}e^{-}$ gamma-ray conversion described by the eq.(\ref{eq:1.3}). The coupling of a hadron $A$ with a virtual photon $\gamma^{*}$ is described by an electromagnetic form factor $F_{A}$. According to this factorization scheme, the differential decay rate can be written as the product of the virtual gamma-ray conversion rate and the decay width $\Gamma(AX\gamma^{*})$\cite{5}:      
\begin{equation}
{\frac{d\Gamma(AXe^{+}e^{-})}{dM}=\frac{2\alpha}{3\pi M}\Gamma(AX\gamma^{*})}\label{eq:2.9}
\end{equation}
Three such processes, for the neutral axial-vectors ($\pi^{0}$, $\eta$) and the vector $\omega$ meson summarized in Table.I. The differential decay width $A\rightarrow Xe^{+}e^{-}$ can be obtained from eq.(\ref{eq:2.9}) with the help of the matrix element: 
\begin{equation}
{\left| M(AX\gamma^{*})\right|^{2}=2p^{2}_{cm}M^{2}_{A}|f_{A}|^{2}}\label{eq:2.10}
\end{equation}
valid also for, $X=\gamma$ and $X=\pi^{0}$\cite{3},\cite{5}:
\begin{figure}[h]
\centering
\includegraphics[width=40mm, height=40mm]{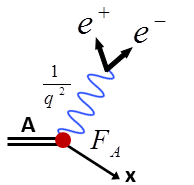}
\caption{A schematic diagram of the meson di-electron Dalitz decay. The blob in the meson-photon vertex describes the meson form factor.}
\end{figure}
\begingroup
\begin{table*}
\centering
\caption{Parametrization of the time-like electromagnetic form factors of $\pi^{0}$, $\eta$ and $\omega$ mesons according to the VDM hypothesis.} 
\begin{tabular}{llr}
\hline
$Dalitz \hspace{2 mm} decay$ \hspace{8 mm}  & $Form\hspace{2 mm} factor \hspace{2 mm} F(M)$ \hspace{8 mm} & $Parameters$ \\
\hline\hline
${\pi^{0}\rightarrow \gamma e^{+}e^{-}}$  & ${F=1+bM^{2}}$ & ${b=5.5}$ ${GeV^{2}}$ \\
 ${\eta\rightarrow \gamma e^{+}e^{-}}$   & ${{\frac{1}{1+(\frac{M}{\Lambda_{\eta}})^{2}}}}$  & ${\Lambda_{\eta}=0.72}$ ${GeV}$ \\
${\omega\rightarrow \pi e^{+}e^{-}}$   & $\frac{\Lambda^{4}_{\omega}}{(\Lambda^{2}_{\omega}-M^{2})^{2}+\Lambda^{2}_{\omega}\gamma^{2}_{\omega}}$  & ${\Lambda_{\omega}=0.65}$ ${GeV}$, ${\gamma_{\omega}=0.04}$ ${GeV}$  \\
\hline
\end{tabular}
\label{table:I}
\end{table*}
\endgroup
\begin{equation}
\begin{split}
{{\left| \Gamma(AX\gamma^{*})\right|^{2}=\frac{p_{cm}}{8\pi M^{2}_{A}} |f_{A}|^{2}}2p^{2}_{cm}M^{2}_{A}}\\ {=\frac{1}{4\pi}p^{3}_{cm}|f_{A}|^{2}}\label{eq:2.11}
\end{split}
\end{equation}
where ${f_{A}(M)}$ is the form factor and $p_{cm}$ is the center mass momentum of the decay ${A\rightarrow X\gamma^{*}}$ products. The latter one is given by the Pauli-K\"all\'en function:
\begin{equation}
{\lambda(x,y,z)=x^{2}+y^{2}+z^{2}-2(xy+xz+yz)}\label{eq:2.12}
\end{equation}
and 
\begin{equation}
{p_{cm}=\frac{\sqrt[]{\lambda}(M^{2}_{A},M^{2}_{X},M^{2})}{2M_{A}}}\label{eq:2.13}
\end{equation}
The electromagnetic form factor defined as ${F_{A}=\frac{f_{A}(M)}{f_{A}(0)}}$ can be introduced by normalization of eq.(\ref{eq:2.11}) to the decay width into the real photon ${A\rightarrow X\gamma}$ obtained with help of eq.(\ref{eq:2.11}):
\begin{equation}
{\Gamma(AX \gamma)=\frac{|f_{A}(0)|^{2}}{4\pi}\frac{\lambda^{3/2}(M^{2}_{A},M^{2}_{X},M^{2})}{8M^{3}_{A}}}\label{eq:2.14}
\end{equation}
and 
\begin{equation}
\begin{split}
{\Gamma(AX \gamma^{*})=}\\ {\Gamma(AX \gamma)\frac{\lambda^{3/2}(M^{2}_{A},M^{2}_{X},M^{2})}{\lambda^{3/2}(M^{2}_{A},M^{2}_{X},0)}|F_{A}(M)|^{2}}\label{eq:2.15}
\end{split}
\end{equation}
The respective electromagnetic form factors $F_{A}$ are collected in Table.I according to the VDM parametrization\cite{8}. The di-electron invariant mass distribution is given by the differential decay ratio distribution:
\begin{equation}
{\frac{dP_{e^{+}e{-}}}{dM}=\frac{1}{\Gamma _{total}}\frac{d\Gamma (AXe^{+}e^{-})}{dM}}\label{eq:2.16}
\end{equation}
The di-electron invariant mass distribution for heavier mesons $\omega$ and $\eta^{'}$ shows a yield enhancement at larger invariant masses. This is due to the $\rho$ meson dominance in the $\gamma\rightarrow e^{+}e^{-}$ conversion process. One should note, that the parametrization of the pion form factor given in Table.I is nothing else but the $F_{\pi}$ given by eq.(\ref{eq:2.8}) but for $M\prec m_{\pi^{0}}$ accessible in the $\pi^{0}$ Dalitz decay. The first two decays from Table.I, agree with the VDM predictions but for the $\omega$ decay a strong discrepancy is observed\cite{2},\cite{8}.
\vspace{1 mm}
\section{Conclusions}
 It well known today that the di-electron Dalitz processes as well for mesons as for baryons are of large interest for the understanding of the structure of hadrons. Furthermore, the pion annihilation, the vector mesons two-body decays and the Dalitz meson decay are important processes in di-electron spectroscopy performed in many experiments such as HADES collaboration and in future CBM collaboration at the GSI in Germany.

\section{Acknowledgments}
The author would like to thank the University of Cyprus. This work done supported by the program: "Studying nuclear matter under extreme conditions of high temperature and high baryonic density".

\end{document}